\documentstyle[12pt]{article}
\begin{document}
\newtheorem{lemma}{Lemma}
\newtheorem{thm}{Theorem}
\newtheorem{cor}{Corollary}
\newtheorem{defn}{Definition}
\rightline{chao-dyn/9711010}

\centerline{\Large{\bf H\"older exponents of irregular signals}} 
\centerline{\Large{\bf and local fractional derivatives }}
\vspace{10pt}

\centerline{ Kiran M. Kolwankar\footnote{email: kirkol@physics.unipune.ernet.in} 
and Anil D. Gangal\footnote{email: adg@physics.unipune.ernet.in}}
\vspace{3pt}

\centerline
{\it Department of Physics, University of Pune, Pune 411 007, India.}

\vspace{6pt}
\begin{abstract}
It has been recognized recently that fractional calculus is useful
for handling scaling structures and processes. We begin this survey 
by pointing out the relevance of the subject to physical situations.
Then the essential definitions and formulae  
from fractional calculus are summarized
and their immediate use in the study of scaling in physical systems is given.
This is followed by a brief summary of classical results. The main
theme of the review rests on the notion of local fractional derivatives. 
There
is a direct connection between local fractional differentiability properties and
the dimensions/ local H\"older exponents of nowhere differentiable functions.
It is argued that local fractional derivatives provide a powerful tool to 
analyse the pointwise behaviour of irregular signals and functions.
\end{abstract}

\vspace{4pt}




\section{Introduction}
In the 1880's, contrary to the notion existing till then that 
a continuous function must be differentiable {\it at least at some  point},
 examples were constructed~\cite{1} to demonstrate
 that it is possible for 
 continuous functions to be {\it no where} differentiable.  Weierstrass's
construction provided
 one such early example. For about three decades after the construction of 
such functions, they  
were still considered to be rather pathological cases 
without any practical relevance to the physical world. Perrin was the first to point out their
application in a real  physical situation, viz.,
 the problem of Brownian motion. 
He also prophesied~\cite{2} the possibility of a simpler way of handling these
functions than differentiable ones.

Recent developments in nonlinear and
nonequilibrium phenomena suggest that such irregular functions occur much
more naturally and abundantly   in formulations of physical theories.
The work on Brownian motion (see~\cite{10}) showed that the graphs of projections of
Brownian paths 
are nowhere differentiable and have
 dimension $3/2$. A generalization of Brownian motion, called fractional 
Brownian motion, ~\cite{2,10}, is known to  
give rise to graphs having dimension between 1 and 2.
It was also observed that 
typical Feynmann paths~\cite{30,31}, like the Brownian paths, are continuous but nowhere 
differentiable.
In fluid systems, passive scalars advected by a turbulent fluid have been shown
\cite{19,20} to have isoscalar surfaces which are highly irregular, in the limit
of diffusion constant going to zero.
Also there exist situations in which  one has to solve a partial
differential equation subject to fractal boundary
conditions, such as Laplace equation near a fractal conducting surface.
For instance, as noted in reference~\cite{39} irregular boundaries may 
appear  to be non-differentiable everywhere upto
a certain spatial resolution, and/or
may exhibit convolutions over many length scales.
 Nowhere differentiable functions can be used to model these boundaries.

In the study of dynamical systems theory, attractors of some systems
are found to be continuous but nowhere differentiable~\cite{15}.
We will consider one specific example~\cite{15,59}
of the dynamical system which gives 
rise to such an attracter. Consider the following map.
\begin{eqnarray}
x_{n+1}&=&2x_n+y_n\;\;\;\;\mbox{mod}\;1, \nonumber\\
y_{n+1}&=&x_n+y_n\;\;\;\;\;\;\mbox{mod}\;1,\\
z_{n+1}&=&\lambda z_n + \cos(2\pi x_n) \nonumber
\end{eqnarray}
where $x$ and $y$ are taken mod 1 and $z$ can be any real number.
In order to keep $z$ bounded, $\lambda$ is chosen between 0 and 1.
The equations for $x$ and $y$ are independent of $z$.  The $x-y$
dynamics are chaotic and are unaffected by the value of $z$.
It was shown in~\cite{15} that if $\lambda > 2/(3+\sqrt{5})$,
the attracter of this map is nowhere differentiable torus. One more example
 in 2-dim phase space is given in section \ref{wrts}.

All these irregular functions are best characterized locally by a {\it H\"older exponent}. We will use following general definition of the H\"older
exponent $h(y)$ which has been used by
various authors~\cite{27,24} recently.
The exponent $h(y)$ of a function $f$
at $y$ is given by $h$ 
such that there exists a polynomial
$P_n(x)$ of order $n$, where $n$ is the largest integer smaller than
$h$, that satisfies
\begin{eqnarray}
\vert f(x) - P_n(x-y) \vert = O(\vert x-y \vert^{h}),
\end{eqnarray}
for $x$ in the neighbourhood of $y$.
This definition serves to classify the behavior of the function at $y$. 

Multifractal measures have been object of many investigations
\cite{26,32,33,34,35,40,69,74}. This 
formalism has met with many applications. Its importance also stems
from the fact that such measures are natural measures to be used in the
analysis of many phenomenon~\cite{36,37}. 
The first mathematical rigorous results concerning multifractals were
given in~\cite{72}.
It may however happen that the object
one wants to understand is a function (e.g. a fractal or multifractal signal)
rather than a set or a measure. For instance one would like to
characterize the velocity field of fully developed turbulence.
Studies on fluid turbulence have shown existence of multifractality  
 in velocity fields of a turbulent fluid at low viscosity~\cite{26}.

In the case of functions having same H\"older exponent $h$,
with $0<h<1$, at every 
point ( Weierstrass function (section \ref{wrts}) is  such a example)
 it is well known~\cite{10} that the
box dimension of its graph is $2-h$. On the other hand, there are functions 
$f(x)$ which do not have constant exponent $h$
at every point but have a range of such exponents.
The set of points having same exponents $\{x\;\vert\; h(x)=h \}$ for such a function
may even constitute a fractal set.
In such situations corresponding functions $f(x)$ are multifractal.

In the light of these findings there has been a renewed interest, 
chiefly among mathematicians, in studying
pointwise behaviour  of multifractal functions.
New tools have been developed to study local behaviour of functions,
which will be of practical use.
Notable among them is the use of Wavelet transforms~\cite{29,38,25,51} which
 have been used with some success to this effect. This
 transform allows one to study the behaviour of functions at different scales
and hence it is  popularly known as a mathematical microscope.
Some well-known functions have been reanalyzed 
using these techniques~\cite{47,48,49,52} 
and shown to possess a spectrum of H\"older exponents, thereby 
establishing their multifractality.

In this article we review a more direct method, which uses
 fractional calculus formalism, to characterize
local behaviour of functions.
The fractional calculus formalism~\cite{7,8,9} allows taking derivatives of fractional order. However, as pointed out in section \ref{secintro} below,
there is a multiplicity of definitions of fractional derivatives.
Hence it is important to recognize the appropriate one which is
suitable for extracting the scaling behaviour.
Recently it was realized~\cite{41} that the concept of local fractional
derivative, introduced in section \ref{secdefn} below
in a more general setting, is ideally
suited for this purpose and that
there is a direct quantitative
connection between the lack of differentiability of a fractal function and 
the dimension of its graph.
Further it provides
 an alternative way of characterizing the local H\"older exponent.
 In this formalism, unlike in wavelet transform,
H\"older exponent shows up naturally since the magnitude of
{\it local fractional derivative} (LFD) jumps 
when its order crosses a {\it critical order}.   
The main theme of this review is centered around this developement.

The organization of the paper is as follows.
In section II we present a brief review of fractional calculus formalism with its recent applications.
The next section discusses fractional differentiability using Weyl's definition
and certain classical results.
In the fourth section we present our definition of LFD
in more general setting and its connection to generalized Taylor series. 
 In section V we 
apply this definition to a specific example, viz, Weierstrass' nowhere 
differentiable function. It is shown that this function, at every point, is
{\it locally fractionally differentiable} for  all orders below $2-s$,
 but not so for orders between $2-s$ and 1, where $s$, $1<s<2$, 
is the box dimension of the graph of the function. In the next section a
general result has been proved, 
 showing relation between local fractional differentiability
 and local H\"older exponent. In section V we demonstrate use of
 LFD in unmasking isolated singularities and
in the study of pointwise behavior of multifractal functions.
Section VI  concludes the article,  pointing out a few possible consequencs of
our results.

\section{Fractional calculus and scaling phenomenon}\label{secintro}
Fractional calculus~\cite{7,8,9} is a study which deals  with generalization 
of differentiation and integration to fractional orders.
There are number of ways (not neccesarily equivalent) of
 defining fractional derivatives and integrations.
We list here some of them, which will be used in this work.
We begin by recalling the Riemann-Liouville definition of fractional integral
of a real function, which is
given by~\cite{7,9}
\begin{eqnarray}
{{d^qf(x)}\over{[d(x-a)]^q}}={1\over\Gamma(-q)}{\int_a^x{{f(y)}\over{(x-y)^{q+1}}}}dy
\;\;\;{\rm for}\;\;\;q<0,\label{def1}
\end{eqnarray}
where the lower limit $a$ is some real number and of the fractional derivative
\begin{eqnarray}
{{d^qf(x)}\over{[d(x-a)]^q}}={1\over\Gamma(n-q)}{d^n\over{dx^n}}{\int_a^x{{f(y)}\over{(x-y)^{q-n+1}}}}dy
\;\;\;{\rm for}\;\;\; n-1<q<n.\label{def2}
\end{eqnarray}
Fractional derivative of a simple function $f(x)=x^p\;\;\;p>-1$ 
is given by~\cite{7,9}
\begin{eqnarray}
{{d^qx^p}\over {d x^q}} = {\Gamma(p+1) \over {\Gamma(p-q+1)}} x^{p-q}\;\;\;
{\rm for}\;\;\;p>-1.\label{xp}
\end{eqnarray}
For future use we note the following properties of fractional derivatives. For $0<q<1$, 
\begin{eqnarray}
{{d^qsin(x)}\over {d x^q}}={{d^{q-1}cos(x)}\over{d x^{q-1}}} \label{sin}
\end{eqnarray}
and
\begin{eqnarray}
{{d^qcos(x)}\over {d x^q}}={x^{-q}\over\Gamma(1-q)}-
{{d^{q-1}sin(x)}\over{d x^{q-1}}} \label{cos}.
\end{eqnarray}
Further the fractional derivative has the property (see ref~\cite{7}), viz,
\begin{eqnarray}
{d^qf(\beta x)\over{d x^q}}={{\beta}^q{d^qf(\beta x)\over{d(\beta x)^q}}}
\end{eqnarray}
which makes it suitable for the study of scaling.
Note the nonlocal character of the fractional derivative and integral
in the definitions equations (\ref{def1}) and (\ref{def2}). Also it is clear
from equation~(\ref{xp}) that unlike in the case of integer derivatives
the fractional derivative of a constant is not zero in general.

Since the definition of fractional derivative contains a lower limit `$a$',
it is natural that different definitions of a fractional differentiability
would arise depending on the lower limit one chooses.
Weyl defined fractional derivatives, with the 
arbitrary limit $a$ in (\ref{def2})
going to $-\infty$, as follows.
\begin{eqnarray}
D^qf(x)={1\over\Gamma(n-q)}{d^n\over{dx^n}}{\int_{-\infty}
^x{{f(y)}\over{(x-y)^{q-n+1}}}}dy
\;\;\;{\rm for}\;\;\; n-1<q<n.\label{def3}
\end{eqnarray}
The merit of this definition is that the
fractional derivative of a periodic 
function according to this definition is a periodic function, which is why
it is suitable in harmonic analysis.

\subsection{Applications to the study scaling in physical systems}
Recent work shows that the fractional calculus formalism is useful
in dealing with scaling phenomena. The purpose of this subsection is to point
out  a few relevant parallel developments even though they are not the part of
the central theme of this paper.

Hilfer~\cite{17,18,70} has generalized the Eherenfest's classification
of phase transition using the fractional derivative.  
Some recent papers~\cite{3,4,5,6} indicate a connection 
between fractional calculus
and fractal structure~\cite{2,10} or fractal processes~\cite{11,22,23}.
 W. G. Gl\"ockle and T. F. Nonnenmacher~\cite{22} have formulated fractional
differential equations for some relaxation processes which are essentially
fractal time~\cite{23} processes.

\subsubsection*{Fractional Brownian Motion}
 Mandelbrot and Van Ness~\cite{11} have 
used fractional integrals to formulate fractal processes such as fractional
Brownian motion.
The fractional Brownian motion is described by the probability distribution
 $B_H(t)$ defined by, 
\begin{eqnarray}
B_H(t)-B_H(0) = {1\over\Gamma(H+1)} \int_{-\infty}^t K(t-t') dB(t'),\;\;\;
0<H<1,
\end{eqnarray}
where $B(t)$ is an ordinary Gaussian random process with average zero and
unit variance and $K(t-t')$ 
\begin{eqnarray}
K(t-t') = \left\{ 
\begin{array}{ll}
(t-t')^{H-1/2} & 0\leq t' <t \\
\{(t-t')^{H-1/2}-(-t)^{H-1/2}\} & t' < 0.
\end{array}
\right.
\end{eqnarray}
The use of the fractional integral kernel is apparent.
 Recently  Sebastian~\cite{55} has given a path integral
representation for fractional Brownian motion whose measure has fractional
derivatives of paths in it.

\subsubsection*{Fractional equations for a class of L\'evy type probability
densities}
L\'evy flights~\cite{64} and L\'evy walks~\cite{65,66} 
have found applications in various branches
of physics \cite{71}, for example in fluid dynamics~\cite{67} and polymers~\cite{68}. 
In~\cite{3} T. F. Nonnenmacher considered a class of normalized one
sided L\'evy type probability densities, which provides 
the length of a jump of a random walker, given by
\begin{eqnarray}
f(x)={a^{\mu}\over{\Gamma(\mu)}} x^{-\mu-1} \exp(-a/x) \;\;\;a>0,\;x>0.
\end{eqnarray}
It is clear that for large $x$
\begin{eqnarray}
f(x) \sim x^{-\mu-1}  \;\;\;\mu>0,\;x>0.
\end{eqnarray}
where $\mu$ is the L\'evy index. In~\cite{3} it was
 shown that these L\'evy-type
probability densities satisfy a fractional integral equation
\begin{eqnarray}
x^{2q}f(x) = a^q {d^{-q}f(x)\over{dx^{-q}}}
\end{eqnarray}
or equivalently a fractional differential equation given by
\begin{eqnarray}
{d^{n-\mu}f(x)\over{dx^{n-\mu}}} = a^{-\mu}{d^n\over{dx^n}}(x^{2\mu}f(x)).
\end{eqnarray}
Here, $n=1$ if $0<\mu<1$ and $n=2$ if $1<\mu<2$, etc. 
An interesting observation of this work is that the L\'evy index is related
to the order of the fractional integral or differential operator.

\subsubsection*{Fractional equations for physical processes}
Modifications of equations governing physical processes such as
 diffusion equation, wave equation and Fokker-Plank equation have been
suggested~\cite{4,5,60,53,54} which incorporate fractional derivatives 
 with respect to time.
In refs.~\cite{4,5} a fractional diffusion equation has been
proposed for the diffusion on fractals. Asymptotic solution of this
equation coincides with the result obtained numerically.
Fogedby~\cite{63} has considered a generalization of the
Fokker-Plank equation involving addition of 
a fractional gradient operator (defined as the Fourier
transform of $-k^{\mu}$) to the usual Fokker-Plank equation
 and performed a renormalization group analysis.

\subsubsection*{Transport in chaotic Hamiltonian systems}
Recently Zaslavasky~\cite{46} showed that the Hamiltonian chaotic dynamics
of particles can be described by a fractional generalization of the
Fokker-Plank-Kolmogorov equation which is defined by two fractional
critical exponents $(\alpha , \beta)$ responsible for the space and
time derivatives of the distribution function correspondingly.
With certain assumptions, the following equation  was derived for the
transition probabilities $P(x,t)$:
\begin{eqnarray}
{\partial^{\beta}P(x,t)\over{\partial t^{\beta}}} = {\partial^{\alpha}\over
{\partial(-x)^{\alpha}}}(A(x)P(x,t)) + {1\over 2} {\partial^{2\alpha}\over
{\partial(-x)^{2\alpha}}}(B(x)P(x,t))
\end{eqnarray}
where 
\begin{eqnarray}
A(x)=\lim_{\Delta t \rightarrow 0} {A'(x;\Delta t)\over{(\Delta t)^{\beta}}}
\end{eqnarray}
and
\begin{eqnarray}
B(x)=\lim_{\Delta t \rightarrow 0} {B'(x;\Delta t)\over{(\Delta t)^{\beta}}}
\end{eqnarray}
with $A'$ and $B'$ being $\alpha^{th}$ and $2\alpha^{th}$ moments respectively.
The exponents $\alpha$ and $\beta$ have been related to anomalous transport
exponent.

We remark  that  most of the applications  reviewed in this section
deal with asymptotic scaling only.

\section{Weyl fractional differentiability and H\"older classes}\label{secweyl}
The purpose of this section is to review the classical work of Stein, Zygmund and others 
\cite{42,43,44,45}
which uses Weyl fractional calculus in the analysis of irregular functions.
Let $f(x)$ be defined in a closed interval I, and let
\begin{eqnarray}
\omega(\delta)=\omega(\delta : f) = \sup \vert f(x)-f(y) \vert
\;\;\;{\rm for }\;\;\;x,y \in I \;\;\;\vert x-y \vert \leq \delta
\end{eqnarray}
The function $\omega(\delta)$ is called the modulus of continuity of $f$.
If for some $\alpha > 0$ we have $\omega(\delta) \leq C {\delta}^{\alpha}$
with $C$ independent of $\delta$, then $f$ is said to satisfy H\"older 
(In old literature it is called Lipschitz)
condition of order $\alpha$. These functions define a class of function
${\Lambda}_{\alpha}$. Notice that with this definition of the modulus of continuity
only the case $0<\alpha \leq 1$ is interesting, since
if $\alpha>1$, $f'(x)$ is zero everywhere and the function is constant.
Hence this definition has been generalized (see~\cite{51}) 
to one where the case $\alpha>1$ is also nontrivial.
The idea is to subtract an appropriate polynomial from a function $f$ at $y$. 
\begin{eqnarray}
\tilde{\omega}(\delta)=\tilde{\omega}(\delta : f) = \sup \vert f(x)-P(x-y) \vert
\;\;\;{\rm for }\;\;\;x,y \in I \;\;\;\vert x-y \vert \leq \delta
\end{eqnarray}
where $P$ is the only polynomial of smallest degree which gives the 
smallest order of magnitude for $\tilde{\omega}$.
Recently, this definition has been widely used~\cite{27,24} to characterize
 the velocity field of a turbulent fluid. 

Following Welland~\cite{45} we introduce
\begin{defn}
$f$ is said to have an $\alpha$th derivative, where $0\leq k < \alpha < k+1$
with integer $k$,
 if $D^{-\beta}f$, $\beta = k+1-\alpha$, has a $k+1$ Peano
derivatives at $x_0$ i.e. there exists a polynomial $P_{x_0}(t)$ of 
degree $\leq k+1$ s. t.
\begin{eqnarray}
(D^{-\beta}f)(x_0+t)-P_{x_0}(t) = o(\vert t \vert^{k+1}) \;\;\;t\rightarrow 0.
\end{eqnarray}
Further if 
\begin{eqnarray}
\{{1\over \rho}\int_{-\rho}^{\rho} \vert (D^{-\beta}f)(x_0+t)-P_{x_0}(t) \vert
^p dt \}^{1/ p} = o(\rho^{k+1}) \;\;\; \rho \rightarrow 0 \;\;\;(1\leq p
< \infty)
\end{eqnarray}
$f$ is said to have an $\alpha$th derivative in the $L^p$ sense.
The $D^{-\beta}$ is in the Weyl sense.
\end{defn}
It is clear that this definition of fractional differentiability is not
local. Particularly the behavior of function at $-\infty$ also plays a crucial
role. The main classical results can be stated using this notion of
differentiability and involve the classes $\Lambda^p_{\alpha}$ and
$N^p_{\alpha}$ which are given by the following definitons.
\begin{defn}
If there exists a polynomial $Q_{x_0}(t)$ of degree $\leq k$ s. t.
$f(x_0+t)-Q_{x_0}(t) = O(\vert t \vert^{\alpha})$ as $t\rightarrow 0$
then $f$ is said to satisfy the condition $\Lambda_{\alpha}$ and if
\begin{eqnarray}
\{{1\over \rho}\int_{-\rho}^{\rho} \vert f(x_0+t)-Q_{x_0}(t) \vert
^p dt \}^{1/ p} = O(\rho^{\alpha}) ,\;\;\; \rho \rightarrow 0 \;\;\;(1\leq p
< \infty)
\end{eqnarray}
$f$ is said to satisfy the condition $\Lambda_{\alpha}^p$.
\end{defn}
\begin{defn}
$f$ is said to satisfy the condition $N_{\alpha}^p$ if for some $\rho > 0$
\begin{eqnarray}
{1\over \rho}\int_{-\rho}^{\rho} {{\vert f(x_0+t)-Q_{x_0}(t) \vert
^p}\over{\vert t \vert^{1+p\alpha}}} dt < \infty.
\end{eqnarray}
\end{defn}
We are now in a position to state the classical results in the form of 
theorems 1 to 4.
Next two theorems~\cite{44,45} give the condition under which the fractional derivative of
a function exists.
\begin{thm}
Suppose that $f$ satisfies the condition $\Lambda_{\alpha}$ at every point 
of a set $E$ of positive measure. Then $D^{\alpha}f(x)$ exists almost everywhere
in $E$ if and only if $f$ satisfies condition  $N_{\alpha}^2$ almost everywhere
in $E$.
\end{thm}
\begin{thm}
The necessary and sufficient condition that $f$ satisfies the condiiton
$N_{\alpha}$ almost everywhere in a set $E$ is that $f$ satisfies the condition
$\Lambda^2_{\alpha}$ and $D^{\alpha}f$ exisits in the $L^2$ sense, almost
everywhere in this set.
\end{thm}
The following results in~\cite{42} tell us how the class of a
function changes when an operation of fractional differentiation is performed.
\begin{thm}
Let $0\leq \alpha < 1$, $\beta > 0$ and suppose that $f\in \Lambda_{\alpha}$.
Then  $D^{-\beta}f \in \Lambda_{\alpha+\beta}$ if $\alpha+\beta < 1$
\end{thm}
\begin{thm}
Let $0<\gamma < \alpha < 1$,.
Then  $D^{\gamma}f \in \Lambda_{\alpha-\gamma}$ if $f\in \Lambda_{\alpha}$
\end{thm}

Though they have their own value, these results are not really adequate
to obtain information regarding irregular behaviour of functions and 
H\"older exponents. We observe that the Weyl definition involves highly
nonlocal information and hence is somewhat unsuitable for treatment of
local scaling behaviour. In the next section we introduce a more
appropriate definition.

\section{Local Fractional Differentiability} \label{secdefn}
 
In our previous work~\cite{41} we introduced the notion of local fractional
derivative and demonstrated its use in the study of local scaling
behaviour. We now briefly explain this notion and use it in
 subsequent sections.

\subsection{Local fractional derivative and critical order}
Recall the observations made in section \ref{secintro}, viz,
(1)~nonlocal information contained in fractional derivative and
(2)~the fractional derivative of a constant is not zero.
The appropriate new notion of differentiability must bypass the
hindrance due to these two properties.
 These difficulties  can be remedied by introducing
\begin{defn} 
If, for a function $f:[0,1]\rightarrow I\!\!R$, the  limit 
\begin{eqnarray}
I\!\!D^qf(y) = 
{\lim_{x\rightarrow y} {{d^q(f(x)-f(y))}\over{d(x-y)^q}}}\label{defloc}
\end{eqnarray}
exists and is finite, then we say that the {\it local fractional derivative} (LFD) 
of order $q$ $(0<q<1)$, at $x=y$, 
exists. 
\end{defn}
In the above definition the lower limit $y$ is treated as a constant.
The subtraction of $f(y)$ corrects for the fact that the fractional
derivative of a constant is not zero. Whereas the limit as $x\rightarrow y$
is taken to remove the nonlocal content.
Advantage of defining local fractional derivative in this manner lies
in its local nature and hence allowing the study of pointwise behaviour
of functions. This will be seen more clearly in section \ref{sectlr}
after the development of Taylor series. 
\begin{defn}
We define {\it critical order} $\alpha$, at $y$, as
$$
\alpha(y) = Sup \{q \;\vert\; {\rm {all\;LFDs\; of\; order\; less\; than\;}} q{{\rm\; exist\; at}\;y}\}. 
$$
\end{defn}
 Though these definitions are interesting only when the critical
order is less than one, for the same reason as that for the first definition
(section \ref{secweyl}) of modulus of continuity
 we extend them for all values of $\alpha > 0$. 
\begin{defn} 
If, for a function $f:[0,1]\rightarrow I\!\!R$, the  limit 
\begin{eqnarray}
I\!\!D^qf(y) =  {\lim_{x\rightarrow y}}
{{d^q(f(x)-\sum_{n=0}^N{f^{(n)}(y)\over\Gamma(n+1)}(x-y)^n)}
\over{[d(x-y)]^q}} \label{deflocg}
\end{eqnarray}
exists and is finite,
where $N$ is the largest integer for which $N^{th}$ derivative of $f(x)$ at
$y$ exists and is finite, then we say that the {\it local fractional
derivative} (LFD) of order $q$ $(N<q<N+1)$, at $x=y$, 
exists. 
\end{defn}
We consider this as the generalization of the local derivative for order 
greater than one. The definition of the critical order remains the same
since, for $q<1$, (\ref{defloc}) and (\ref{deflocg}) agree.
This definition extends the applicability of LFD to functions where the first
derivative exists but are still irregular due to the nonexistence of some 
higher order derivative.
As an example we note that the critical order of 
$f(x)=a+bx+c\vert x\vert ^{\gamma}$ at origin,
according to definitions 5 and 6 is $\gamma$.

\noindent
{\bf Remark:} 1)It is interesting to note that the same definition of
LFD can be used for negative values of the critical order between -1 and 0.
For this range of critical orders $N=-1$ and the sum in equation 
(\ref{deflocg}) is empty. As a result the expression for LFD becomes
\begin{eqnarray}
I\!\!D^qf(y) = 
{\lim_{x\rightarrow y} {{d^qf(x)}\over{[d(x-y)]^q}}}
\end{eqnarray}
An equivalence between the critical order and the H\"older exponent,
for positive values of critical order, will be proved in section \ref{eql}.
Here we would like to point out that the
negative H\"older exponents do arise in real physical situation of 
turbulant velocity field (see~\cite{50,56} and references therein).

\noindent
2)Another way of generalizing the LFD to the values of critical order
beyond 1 would have been to write it as
\begin{eqnarray}
I\!\!D^qf(y) = 
{\lim_{x\rightarrow y} {{d^q(f^{(N)}(x)-f^{(N)}(y))}\over{[d(x-y)]^q}}}
\end{eqnarray}
But the existence of $N^{th}$ derivative of $f$ at $x$ may not be 
guaranteed in general. Such a situation may arise in the case of 
multifractal functions to be treated in section \ref{mff}.

\subsection{LFD  and generalized Taylor series}\label{sectlr}
In order to see the information contained in the LFD we consider
fractional Taylor's series with a remainder term for a real function $f$.
Let
\begin{eqnarray}
F(y,x-y;q,N) = {d^q(f(x)-\sum_{n=0}^{N}{f^{(n)}(y)\over{\Gamma(n+1)}}(x-y)^n)
\over{[d(x-y)]^q}} \label{F}
\end{eqnarray}
and
\begin{eqnarray}
\widetilde{F_N}(x,y) = f(x)-\sum_{n=0}^N{f^{(n)}(y)\over\Gamma(n+1)}(x-y)^n).
\label{Ftilde}
\end{eqnarray}
It is clear that
\begin{eqnarray}
I\!\!D^qf(y)=F(y,0;q,N)
\end{eqnarray}
Now, for $N<q<N+1$, it can be shown that
\begin{eqnarray}
\widetilde{F_N}(x,y)
&=& {I\!\!D^qf(y)\over \Gamma(q+1)} (x-y)^q \nonumber\\
&&+ {1\over\Gamma(q+1)}\int_0^{x-y} {dF(y,t;q,N)\over{dt}}{(x-y-t)^q}dt\label{taylor}
\end{eqnarray}
i.e.
\begin{eqnarray}
f(x) = \sum_{n=0}^{N}{f^{(n)}(y)\over{\Gamma(n+1)}}(x-y)^n
 + {I\!\!D^qf(y)\over \Gamma(q+1)} (x-y)^q + R_1(x,y) \label{taylor2}
\end{eqnarray}
where $R_1(x,y)$ is a remainder given by
\begin{eqnarray}
R_1(x,y) = {1\over\Gamma(q+1)}\int_0^{x-y} {dF(y,t;q,N)\over{dt}}{(x-y-t)^q}dt
\end{eqnarray}
We note that the local fractional derivative as defined above
(not just fractional derivative), along with the first $N$ derivatives,
 provides an approximation
of $f(x)$  
in the vicinity of $y$. 
 We further note that the terms
on the RHS of eqn(\ref{taylor}) are nontrivial and finite only in the case $q=\alpha$, the critical order.
In ref.\cite{21} a fractional Taylor
series was constructed by Osler
using usual (not local in the present sense) fractional derivatives. 
His results are, however, applicable to analytic functions and cannot be 
used for non-smooth functions directly. Further Osler's
formulation involves terms with negative orders also and hence is not suitable
for approximating schemes.

When $0<q<1$ we get as a special case
\begin{eqnarray}
f(x) = f(y)
 + {I\!\!D^qf(y)\over \Gamma(q+1)} (x-y)^q + Remainder \label{taylor01}
\end{eqnarray}
provided the RHS exists.
 One may  note in equation(\ref{taylor01}) that when $q$ is set equal to
one in the above approximation one gets
the equation of the tangent. 
It may be recalled that all the curves passing through a point $y$ and having same tangent
form an equivalence class (which is modelled by a linear behavior). 
Analogously all the functions (curves) with the same critical order $\alpha$
and the same $I\!\!D^{\alpha}$
will form an equivalence class modeled by power law $x^{\alpha}$  
This is how one may
generalize the geometric interpretation of derivatives in terms of tangents.  
This observation is useful when one wants to approximate an irregular 
function by a piecewise smooth (scaling) function.

\section{Fractional Differentiability of Weierstrass Function}\label{wrts}
Consider a form of Weierstrass function, viz,
\begin{eqnarray}
W_{\lambda}(t) = \sum_{k=1}^{\infty} {\lambda}^{(s-2)k} sin{\lambda}^kt,\;\;\;\;
\lambda>1,\;\;\;1<s<2,\;\;\;t\;\mbox{real}.\label{Weier}
\end{eqnarray}
Note that $W_{\lambda}(0)=0$.
The box dimension of the graph of this function is $s$. The Hausdorff
dimension of its graph is still unknown. The best known bounds are given
by Mauldin and Williams~\cite{73} where they have shown that there is
a constant $c$ such that
$$
s - {c \over{\log{\lambda}}} \leq dim_Hgraph f \leq s.
$$ 
Recently it was shown~\cite{79} that if a random phase is added to 
each sine term in (\ref{Weier}) then the Hausdorff dimension of the graph
of the resulting function is $s$.
These kind of functions have been studied in detail in~\cite{10,12,13,14}.

We note that there are dynamical systems with graphs of such functions
as invariant sets. For example,
let $g:I\!\!R \rightarrow I\!\!R$ be differentiable, and let  $h:I\!\!R^2 \rightarrow I\!\!R^2$
be given by 
\begin{eqnarray}
h(x,t)=(\lambda t, \lambda^{2-s}(x-g)t)).
\end{eqnarray}
Then the graph of $f$ can easily be seen to be 
a repeller of for $h$, where $f$ is the function given by
\begin{eqnarray}
f(t) = \sum_{k=1}^{\infty} {\lambda}^{(s-2)k} g({\lambda}^kt).
\end{eqnarray}

In the following two subsections we prove lower and upper bound
on critical order of the Weierstrass function.

\subsection{Lower bound on critical order}

To check the fractional differentiability at any point, say $\tau$,
we use $t'=t-\tau$ and $\widetilde{W}_{\lambda} (t',\tau )=
W_{\lambda}(t'+\tau )-W_{\lambda}(\tau)$ so that
$\widetilde{W}_{\lambda}(0,\tau )=0$. We have
\begin{eqnarray}
\widetilde{W}_{\lambda} (t' ,\tau ) &=& \sum_{k=1}^{\infty} {\lambda}^{(s-2)k} sin{\lambda}^k(t' +\tau)-
\sum_{k=1}^{\infty} {\lambda}^{(s-2)k} sin{\lambda}^k\tau \nonumber\\
&=&\sum_{k=1}^{\infty} {\lambda}^{(s-2)k}(cos{\lambda}^k\tau sin{\lambda}^kt' +
sin{\lambda}^k\tau(cos{\lambda}^kt' -1)) \label{c}
\end{eqnarray}
Now we take fractional derivative of this with respect to $t'$.
\begin{eqnarray}
{d^q\widetilde{W}_{\lambda} (t' ,\tau )\over{dt'^q}}=
\sum_{k=1}^{\infty} {\lambda}^{(s-2+q)k}\left(cos{\lambda}^k\tau 
{d^qsin{\lambda}^kt'\over{d({\lambda}^kt')^q}} +
sin{\lambda}^k\tau{d^q(cos{\lambda}^kt' -1)\over{d({\lambda}^kt')^q}}\right) \label{a}
\end{eqnarray}
From equations (\ref{sin}), (\ref{cos}) 
and second mean value theorem it follows that the fractional 
derivatives inside the above sum is 
bounded uniformly for all values of ${\lambda}^kt$. 
This implies that the series on the right 
hand side will converge uniformly for $q<2-s$, justifying our action of taking
the fractional derivative operator inside the sum.

Also as $t' \rightarrow 0$ for
any $k$ the fractional derivatives in the summation of equation (\ref{a}) goes to zero.
Therefore it is easy to see from this that
\begin{eqnarray}
I\!\!D^qW_{\lambda}(\tau) = {\lim_{t'\rightarrow 0} {{d^q\widetilde{W}_{\lambda}
(t',\tau )}\over{dt'^q}}}=0\;\;\;
{\rm for} \;\;\;q<2-s.
\end{eqnarray}
This shows that $q^{th}$ local derivative of the Weierstrass function exists and
is continuous, at any point, for $q<2-s$.

\subsection{Upper bound on critical order}
For $q>2-s$,  right hand side of the equation (\ref{a})  seems to diverge.
We now prove that the LFD of order $q>2-s$ in fact does not
exist.
To do this we write the Weierstrass function as follows.
\begin{eqnarray}
W_{\lambda}(t) = \sum_{k=1}^{N} {\lambda}^{(s-2)k} sin{\lambda}^kt
+ {\lambda}^{(s-2)N}W_{\lambda}({\lambda}^Nt).
\end{eqnarray}
We now write
\begin{eqnarray}
{{d^q(W_{\lambda}(t)-W_{\lambda}(t'))}\over{d(t-t')^q}}  
&=& {\sum_{k=1}^{N} {\lambda}^{(s-2+q)k}{{d^q(sin({\lambda}^kt)-sin({\lambda}^kt'))}\over {d({\lambda}^kt)^q}}}\nonumber\\
&&+ \lambda^{(s-2+q)N}
 {{d^q(W_{\lambda}(\lambda^{N}t)-W_{\lambda}(\lambda^{N}t'))}
\over{d(\lambda^{N}(t-t'))^q}} \label{decomp}
\end{eqnarray}
We choose $N$ such that $\lambda^{-(N+1)} < \vert t-t' \vert \leq \lambda^{-N}$.
Now since $\vert sin({\lambda}^kt)-sin({\lambda}^kt')\vert \leq 
\lambda^k\vert t-t' \vert$
\begin{eqnarray}
{{d^q|sin({\lambda}^kt)-sin({\lambda}^kt')|}\over {d[{\lambda}^k(t-t')]^q}}
\leq C \lambda^{k(1-q)}\vert t-t' \vert^{1-q}
\end{eqnarray}
This implies that the absolute value of the
 first term in equation (\ref{decomp}) is bounded
from above by $C\lambda^{(s-2+q)N}$.
Now since the first derivative of the Weierstrass function does not exist
at any point there exists a sequence of points $t_n$ approaching $t'$
such that $|W_{\lambda}(\lambda^{N}t_n)-W_{\lambda}(\lambda^{N}t')| \geq 
c\lambda^{N}\vert t_n-t' \vert$. Therefore there exists a sequence $t_n$ 
such that
\begin{eqnarray}
{{d^q|W({\lambda}^Nt_n)-W({\lambda}^Nt')|}\over {d[{\lambda}^N(t_n-t')]^q}}
\geq c \lambda^{N(1-q)}\vert t_n-t' \vert^{1-q}
\end{eqnarray}
and this is valid for every $c$ for large enough $n$.
This implies that 
\begin{eqnarray}
\lambda^{(s-2+q)N}
 {{d^q|W_{\lambda}(\lambda^{N}t_n)-W_{\lambda}(\lambda^{N}t')|}
\over{d(\lambda^{N}(t_n-t'))^q}} \geq c \lambda^{(s-2+q)N}
\end{eqnarray}
\begin{eqnarray}
\vert{{d^q(W_{\lambda}(t_n)-W_{\lambda}(t'))}\over{d(t_n-t')^q}} &-&
\lambda^{(s-2+q)N}
 {{d^q(W_{\lambda}(\lambda^{N}t_n)-W_{\lambda}(\lambda^{N}t'))}
\over{d(\lambda^{N}(t_n-t'))^q}}\vert \nonumber\\
&\leq&
{\sum_{k=1}^{N} {\lambda}^{(s-2+q)k}{{d^q|sin({\lambda}^kt_n)-sin({\lambda}^kt')|}\over {d({\lambda}^k(t_n-t'))^q}}}
\end{eqnarray}
Therefore we get
\begin{eqnarray}
\vert{{d^q(W_{\lambda}(t_n)-W_{\lambda}(t'))}\over{d(t_n-t')^q}} -
c\lambda^{(s-2+q)N}
 \vert \leq
C {\lambda}^{(s-2+q)k}
\end{eqnarray}
This implies that
\begin{eqnarray}
\vert{{d^q(W_{\lambda}(t_n)-W_{\lambda}(t'))}\over{d(t_n-t')^q}}\vert \geq
C' {\lambda}^{(s-2+q)k}
\end{eqnarray}
for $C' > 0$.
From this it is clear that the LFD of order greater that $2-s$ does not exist.
This concludes the proof.

Summarizing, therefore, the critical order of the Weierstrass function 
is $2-s$ at all points.
It may be noticed that this proof is valid for any $\lambda > 1$. This
generalizes a similar result of~\cite{10,41} which is valid only for sufficiently
large $\lambda$.
Thus there is a direct connection between dimension and the differentiability
properties for $W_{\lambda}(t)$. As seen below, this observation is not
restricted to $W_{\lambda}(t)$ but is quite general.

\subsection{L\'evy index of a L\'evy flights and critical order}

 Schlesinger et al~\cite{16} have considered a 
L\'evy flight on a one dimensional 
periodic lattice where a particle jumps from one lattice site 
to other  with the probability given by
\begin{eqnarray}
P(x) = {{{\omega}-1}\over{2\omega}} \sum_{j=0}^{\infty}{\omega}^{-j}
[\delta(x, +b^j) + \delta(x, -b^j)]
\end{eqnarray}
where $x$ is magnitude of the jump, $b$ is a lattice spacing and $b>\omega>1$. 
$\delta(x,y)$ is a Kronecker 
delta.
The characteristic function for $P(x)$ is given by
\begin{eqnarray}
\tilde{P}(k) = {{{\omega}-1}\over{2\omega}} \sum_{j=0}^{\infty}{\omega}^{-j}
cos(b^jk).
\end{eqnarray}
which is nothing but the Weierstrass function.
For this distribution the L\'evy index is $\log{\omega}/\log{b}$, which can be
identified as critical order of $\tilde{P}(k)$. 

More generally for the L\'evy distribution with index $\mu$ 
the characteristic function
is given by
\begin{eqnarray}
\tilde{P}(k) =A \exp{c\vert k \vert^{\mu}}.
\end{eqnarray}
Critical order of this function at $k=0$
also turns out to be same as $\mu$. Thus the L\'evy index can be identified as
the critical order of the characteristic function at $k=0$.

\section{Equivalence  between critical order and the H\"older exponent}\label{eql}
We recall the definition of H\"older exponent $h(y)$ of a function $f$
at $y$~\cite{27,24} as  $h$ such that there exists a polynomial
$P_n(x)$ of order $n$, where $n$ is the largest integer smaller than $h$,
 that satisfies
\begin{eqnarray}
\vert f(x) - P_n(x-y) \vert = O(\vert x-y \vert^{h}), \label{holder}
\end{eqnarray}
for $x$ in the neighbourhood of $y$. 
When $h$ is restricted between 0 and 1 the equation (\ref{holder}) takes the
form
\begin{eqnarray}
\vert f(x) - f(y) \vert = O(\vert x-y \vert^{h}), 
\end{eqnarray}
The H\"older exponent characterizes the
behaviour of the function around a given point. One can study the function
at every point and find its pointwise H\"older exponents.
According to equation (\ref{holder}) an analytic function has 
$h=\infty$ at every point. 

As can be seen clearly that the definition of the H\"older exponent is not 
algorithmic  and hence methods need to be developed for its determination.
In~\cite{41} the following result was proved in a slightly different form,
which establishes an equivalence between the H\"older exponent and the
critical order. 
\begin{thm}\label{thm1}
The continuous function $f(x)$ has H\"older exponent $\alpha$,
$0< \alpha < 1$, at $y$
iff $I\!\!D^qf(y) = 0$ for all $q<\alpha$ and $I\!\!D^qf(y)$ does not
exist for $1>q>\alpha$.
\end{thm}
This result can be understood from the simple example of fractional
derivative of a power $x^p$,
given in equation (\ref{xp}). In this equation if $0<p<1$
and $0<q<1$ then the RHS exists. But if we take a limit $x\rightarrow 0$
then we get
\begin{eqnarray}
\lim_{x\rightarrow 0} {d^qx^p\over{dx^q}} =\left\{ \begin{array}{ll}
0 & \mbox{if $q<p$} \\
\mbox{const} & \mbox{if $p=q$}\\
\infty & \mbox{otherwise}
\end{array} \right.
\end{eqnarray}
The LHS in the above equation is nothing but the LFD of $x^p$. This 
shows that the critical order gives the exponent $p$.
This is expected to happen for any function and at any point with a local power law behaviour.

The following corrollary follows from the theorem \ref{thm1}
and a well known result giving relation between H\"older exponent
and box dimension of a graph of a fractal function~\cite{10}.
\begin{cor}
If the critical order of a function $f(x)$ at every point $x$ is $\alpha$
then $dim_Bf = 2 - \alpha$ where $dim_Bf$ is a box dimension of the graph
of the function $f$.
\end{cor}

With a slight modification in the proof of theorem \ref{thm1}
 a general result giving equivalence between the H\"older
exponent and the critical order using the general definition of LFD follows.
The functions $F$ and $\widetilde{F}$ are defined in equations (\ref{F}) and
(\ref{Ftilde}) respectively.

\begin{thm}\label{thm2}
 Let $f:[0,1]\rightarrow I\!\!R$ be a continuous function.

a)If
$I\!\!D^qf(y) = 0$ where $N< q < N+1$,
for some $y$, 
then  $ h(y)\geq q$.

b)If there exists a sequence  $x_n \rightarrow y$ as
$n \rightarrow \infty$ such that
\begin{eqnarray}
\lim_{n\rightarrow \infty} F(y,x_n-y;q,N) =\pm \infty\;\;\;
,\nonumber
\end{eqnarray}
for some $y$, 
then $h(y) \leq q$.
\end{thm}

\noindent

Similarly  a following converse of the above theorem can also be proved.

\begin{thm} \label{thm3}

Let $f:[0,1]\rightarrow I\!\!R$ be a continuous function.

a) Suppose 
\begin{eqnarray}
\vert \widetilde{F}_N(x,y) \vert \leq c\vert x-y \vert ^{\alpha}, \nonumber
\end{eqnarray}
where $c>0$, $N<\alpha <N+1$ and $|x-y|< \delta$ for some $\delta >0$.
Then $I\!\!D^qf(y) = 0$ for any $q < \alpha$
for $y\in (0,1)$

b) Suppose that for $y\in (0,1)$ and for each $\delta >0$ there exists x such that
$|x-y| \leq \delta $ and
\begin{eqnarray}
\vert \widetilde{F}_N(x,y) \vert \geq c{\delta}^{\alpha}, \nonumber
\end{eqnarray}
where $c>0$, $\delta \leq {\delta}_0$ for some ${\delta}_0 >0$ and $0<\alpha<1$.
Then there exists a sequence $x_n \rightarrow y$ as $n\rightarrow \infty$
such that
\begin{eqnarray}
\lim_{n\rightarrow \infty} F(y,x_n-y;q,N)=\pm \infty\;\;\;
{\rm for}\;\; q>\alpha\;\; 
\nonumber
\end{eqnarray}
\end{thm}

These two theorems give an equivalence between H\"older exponent and the critical order of fractional differentiablity. Their proofs are similar to
that of theorem \ref{thm1}.

\section{Isolated masked singularities}
The purpose of this section is to demonstrate the use of LFD to
detect masked singularities. We will consider only isolated singularities.
   We choose the simplest example $f(x)=\sum_{n=0}^N 
a_nx^n + ax^{\alpha},\;\;\;N<\alpha
<N+1,\;\;\;x>0$.  Critical order at $x=0$ gives the order of 
singularity at that point whereas
the value of the LFD $I\!\!D^{q=\alpha}f(0)$, viz 
$a\Gamma(\alpha+1)$, gives strength of the singularity.

Using LFD we can detect a weaker singularity masked by a stronger singularity.
As demonstrated below, we can estimate and subtract the contribution due to 
stronger singularity from the 
function and find out the critical order of the remaining function.
Consider, for example, the function
\begin{eqnarray}
f(x)=\sum_{n=0}^N 
a_nx^n + ax^{\alpha}+ \sum_{n=N+1}^M 
a_nx^n + bx^{\beta},
\label{masked}
\end{eqnarray}
where $N<\alpha <N+1<M<\beta <M+1$ and $x>0$.
LFD of this function at $x=0$ of the order $\alpha$ is 
$I\!\!D^{\alpha}f(0)=a\Gamma(\alpha+1)$.
Using this estimate of stronger singularity we now write
 $$
G(x;\alpha)=f(x)-\sum_{n=0}^N{f^{(n)}(0)\over{\Gamma(n+1)}}x^n
-{I\!\!D^{\alpha}f(0)\over\Gamma(\alpha+1)}x^{\alpha}.
$$
The critical order of this function, at $x=0$, is $\beta$ which
is a masked singularity. 
 Notice that the estimation of the weaker singularity was possible in the
above calculation just because the LFD gave the coefficient of $x^{\alpha}/
{\Gamma(\alpha+1)}$. This suggests that using LFD, one should be able to extract secondary singularity spectrum
masked by the primary singularity spectrum of strong singularities. Hence one 
can gain more insight into the processes giving rise to irregular
behavior. 

Comparison of two methods of studying pointwise behavior 
of functions, one using wavelets and the other using LFD, 
shows that characterisation of H\"older classes of
functions using LFD is direct and involves fewer assumptions. 
Characterisation of H\"older class of functions with oscillating singularity, 
e.g. $f(x)=x^{\alpha}sin(1/x^{\beta})$ ($x>0$, $0< \alpha <1$ and $\beta>0$), 
using wavelets needs two exponents~\cite{25}. 
Using LFD, owing to theorem I and II critical order 
directly  gives the H\"older exponent for such a function.  

It has 
been shown in the context of wavelet transforms that 
one can detect singularities masked by regular polynomial
behavior~\cite{27} by choosing a appropriate analysing wavelet. (Wavelets with 
 first $n$, for some suitable $n$,
moments vanishing are considered appropriate).
 If one has to extend the wavelet method 
to unmask  weaker singularities,
 one would then require analysing wavelets with fractional moments vanishing.
 Notice that
 one may require this condition along with the condition 
on first $n$ moments. Further the class of functions to be analysed is in
general restricted in these analyses. These restrictions essentially arise
from the asymptotic properties of the wavelets used.
 On the other hand, these restrictions are not relevant while using LFD.

\section{Multifractal function}\label{mff}
We saw in section \ref{wrts} that the Weierstrass function 
is a fractal function, i.e., it has the same
H\"older exponent at every point. But there are multifractal functions which have different H\"older exponents at different points. 
These functions can be used to model various intermitent signals arising
in physical applications.
 Since the critical order gives the local 
and pointwise behavior of the function, conclusions of the theorem 
\ref{thm1}, \ref{thm2} and \ref{thm3} will
carry over even to the case of multifractal functions where we have 
different H\"older exponents at different points.
Selfsimilar multifractal functions have been constructed by Jaffard~\cite{24}.
We give one specific example of such a function.
This function is a solution $F$ of the functional equation 
\begin{eqnarray} 
F(x)=\sum_{i=1}^d {\lambda}_iF(S_i^{-1}(x)) + g(x),\;\;\;x\;\mbox{real}.
\end{eqnarray}
where $S_i$'s are the affine transormations of the kind 
$S_i(x)={\mu}_ix+b_i$ (with $\vert \mu_i \vert < 1$ and $b_i$'s real)
 and  $\lambda_i$'s 
are some real numbers and $g$ is any sufficiently smooth function  
(it is assumed that $g$ and its
 derivatives  have fast decay). For the sake of illustration 
we choose ${\mu}_1={\mu}_2=1/3$, $b_1=0$, $b_2=2/3$, 
${\lambda}_1=3^{-\alpha}$, ${\lambda}_2=3^{-\beta}$ ($0<\alpha<\beta<1$) and 
\begin{eqnarray}
g(x)&=& sin(2\pi x)\;\;\;\;\;\;{\rm if}\;\;\;\; x\in [0,1]\nonumber\\
&=&0\;\;\;\;\;\;\;\;\;{\rm otherwise}. \nonumber
\end{eqnarray}
Such functions are studied in detail in~\cite{24} using wavelet transforms
where it was shown that  the  above functional equation (with the
parameters we have chosen)
has a unique solution $F$. Further at any point
$F$ either has H\"older exponents ranging from 
$\alpha$ to $\beta$ or is smooth.  A sequence of points $S_{i_1}(0),\;\;$$ \;S_{i_2}S_{i_1}(0),\;\;$ $ 
\cdots,\;\;\; $$S_{i_n}\cdotp \cdotp \cdotp $$S_{i_1}(0), \;\;\cdots$,
where $i_k$ takes values 1 or 2, 
tends to a point in $[0,1]$ (in fact to a point of a triadic
Cantor set) and for the values of 
${\mu}_i$s we have chosen this correspondence between sequences and limits 
is one to one. 
The solution of the above functional equation is given by~\cite{24} 
\begin{eqnarray}
F(x)=\sum_{n=0}^{\infty}\;\;\;\sum_{i_1,\cdots,i_n=1}^2{\lambda}_{i_1}\cdots{\lambda}_{i_n}
g(S_{i_n}^{-1}\cdots S_{i_1}^{-1}(x)). \label{soln}
\end{eqnarray}
Note that with the above choice of parameters the inner sum in (\ref{soln})
 reduces to a single term. Jaffard~\cite{24} has shown that the local 
H\"older exponent at $y$ is
\begin{eqnarray}
h(y)=\liminf_{n\rightarrow \infty}{{\log{({\lambda}_{{i_1}(y)}\cdots{\lambda}_{{i_n}(y)})}}
\over{\log{({\mu}_{{i_1}(y)}\cdots{\mu}_{{i_n}(y)})}}},
\end{eqnarray}
where $\{i_1(y)\cdot\cdot\cdot i_n(y)\}$ is a sequence of intergers appearing in the
 sum in equation(\ref{soln}) at a point $y$.
 It is clear that $h_{min}=\alpha$ and
$h_{max}=\beta$. The function $F$ at the points of a triadic cantor
 set have $h(x) \in [\alpha , \beta]$
and at other points it is smooth ( where $F$ is as smooth as $g$).
 Benzi et. al.~\cite{28} have constructed multifractal functions which
are random in nature unlike the above nonrandom functions.
For still another approach also see~\cite{80}.

Several well-known `pathological' functions have been reanalyzed 
in~\cite{47,48,49,52} 
and found to have multifractal nature.
Here we consider one example of classical multifractal function.
\begin{eqnarray}
R(x) = \sum_{n=1}^{\infty}{1\over{n^2}}\sin(\pi n^2x)
\end{eqnarray}
This function was proposed by Riemann. It turns out that the regularity
of this function varies strongly from point to point. Hardy and Littlewood
\cite{57} proved that $R(x)$ is not differentiable at $x_0$ if $x_0$
is irrational or if $x_0$ can not be written as $2p+1/2q+1$ ($p,q \in N$).
In fact they showed that the H\"older exponent at these points $\leq 3/4$.
Gerver~\cite{58} proved the differentiability of $R(x)$ at points of the form
$2p+1/2q+1$ ($p,q \in N$). At these points the H\"older exponent is 3/2.
This function has also been studied in~\cite{61,62}.
Jaffard \cite{47} has recently shown that the dimension 
spectrum of the Riemann function
is given as below.
\begin{eqnarray}
d(\alpha) = \left\{ 
\begin{array}{ll}
4\alpha -2 & \mbox{if}\;\;\; \alpha \in [{1\over2},{3\over4}]\\
0 & \mbox{if}\;\;\; \alpha = {3\over2}\\
-\infty & \mbox{otherwise} 
\end{array}   \right.
\end{eqnarray}
where $d(\alpha)$ gives the Hausdorff dimension of the set when the 
H\"older exponent is $\alpha$.

LFD forms one method of studying pointwise behaviour of such multifractal 
functions alongwith other methods and may cosiderably reduce the 
analysis involved. This
fact was demonstrated in~\cite{41} on a specific example of self-similar
multifractal function given by equation (\ref{soln}).

\section{Concluding remarks}
First we reviewed the applications of various fractional differential
equations in different physical situations. It was noted that most
of these applications dealt with the asymptotic scaling. Further we reviewed
the classical results within the framework of Weyl fractional calculus.
However these results were found to be
 inadequate for the study of pointwise behavior
of fractal and multifracal functions. The notion of LFD as developed in
\cite{41} was found suitable for this purpose. In terms of the LFD it
was also possible to write a fractional Taylor series of a function
(useful in an analytic treatment of approximations). It was pointed out that
generalization of the notion of tangents to the graph of a function
(useful for geometric purposes) is also possible using LFD.

It was established that  the critical order of the
Weierstrass function is related to the box dimension of its graph. If  the dimension of
the graph of such a function is $1+\gamma$, the critical order is $1-\gamma$. When 
$\gamma$ approaches unity the function becomes increasingly irregular and local fractional
differentiability is lost accordingly. Thus there is a direct quantitative connection between the 
dimension of the graph and the fractional differentiability property of the function.
This is one of the remarkable conclusions of the new approach.
An important consequence of this approach
 is that a classification of continuous paths 
(e.g. fractional Brownian paths) or
functions according to local fractional differentiablity properties is also
a classification according to  dimensions (or H\"older exponents).

Also  the L\'evy index of a L\'evy flight on one dimensional 
lattice is identified as
the critical order of the characteristic function of the walk. More generally,
the L\'evy index of a L\'evy  distribution is identified as
the critical order of its characteristic function at the origin.

We have argued and demostrated that LFDs are useful for studying isolated singularities and singularities masked by the stronger singularity (not just by
regular behavior). It was also shown that the pointwise
behavior of irregular (fractal or multifractal) functions can be studied
using the methods of this paper.

We note, however, that the treatment of random irregular 
functions as well as multivariable irregular functions 
 is badly needed. We hope that these problems can be tackled in near future.

\section*{Acknowledgements}

One of the authors (KMK) is grateful to CSIR (India) for financial assistance and the other author
(ADG) is grateful to UGC (India) for financial assistance during initial stages of the work. We thank the referee and K. R. Srinivasan for pointing out some
missing references. In particular the refree pointed our attention to 
the mathematical work~\cite{75,76} dealing with multifractal measures. Referee
also informed us about the recent work by M. Z\"ahle~\cite{77,78}.



\begin{thebibliography}{99}
\bibitem{1} T. W. K\"orner  {\it Fourier Analysis} ( Cambridge University Press, Cambridge, 1989)
\bibitem{81} For historical remarks and construction see,
for example, \cite{1,2,10}.
\bibitem{2} B. B. Mandelbrot   {\it The Fractal Geometry of Nature}
( Freeman, New York, 1977).
\bibitem{10} K. Falconer  {\it Fractal Geometry} ( John Wiley, New York, 1990).
\bibitem{30} R. P. Feynmann and A. R. Hibbs {\it Quantum Mechanics
and Path Integrals} ( McGraw-Hill, New York, 1965). 
\bibitem{31} L. F. Abott and M. B. Wise {\it Am. J. Phys.} {\bf 49}, 37 (1981).
\bibitem{19} P. Constantin, I. Procaccia and K. R. Sreenivasan {\it Phys. Rev. Lett.} {\bf 67}, 1739 (1991).
\bibitem{20} P. Constantin and I. Procaccia {\it Nonlinearity} {\bf 7},
 1045 (1994).
\bibitem{39} K. Sarkar and C. Meneveau {\it Phys. Rev. E} {\bf 47}, 957 (1993).
\bibitem{15} J. L. Kaplan, J. Malet-Peret and J. A. Yorke {\it Ergodic Th. and Dyn. Syst.}{\bf 4}, 261 (1984).
\bibitem{59} J. D. Farmer, E. Ott and J. A . Yorke Physica D {\bf 7}, 153
(1983).
\bibitem{27} J. F. Muzy, E. Bacry and A. Arneodo {\it Phys. Rev. E} {\bf 47}, 
875 (1993).
\bibitem{24} S. Jaffard, To appear in {\it SIAM J. of Math. Anal.}.
\bibitem{26} U. Frisch and G. Parisi, in {\it Turbulence and Predictability in Geophysical Fluid
Dynamics and Climate Dynamics} edited by M Ghil, R. Benzi and G. Parisi 
(North-Holland, Amsterdam, 1985).
\bibitem{32} R. Benzi, G. Paladin, G. Parisi  and A. Vulpiani 
{\it J. Phys. A} {\bf 17}, 3521 (1984).
\bibitem{33} T. C. Halsey, M. H. Jensen,  L. P. Kadanoff, I. Procaccia and B. I. Shraiman {\it Phys. Rev. A} {\bf 33}, 1141 (1986).
\bibitem{34} P. Collet, J. Lobowitz and A. Porzio 
{\it J. Stat. Phys.}
{\bf 47}, 609 (1987).
\bibitem{35} M. H. Jensen, L. P. Kadanoff and I. Procaccia 
{\it Phys. Rev. A} {\bf 46}, 1409 (1987).
\bibitem{40} B. B. Mandelbrot {\it Pure Appl. Geophys.} {\bf 131}, 5 (1989).
\bibitem{69} C. Meneveau and K. R. Sreenivasan, {\it Phys. Rev. Lett.}
{\bf 59}, 1424 (1987).
\bibitem{74} A. B. Chhabra and K. R. Srinivasan, {\it Phys. Rev. Lett.}
{\bf 68} 2762 (1992).
\bibitem{36} J. Feder, {\it Fractals} (Pergamon, New York, 1988).
\bibitem{37} T. Vicsek, {\it Fractal Growth Phenomenon} 
(World Scientific, Singapore, 1989).
\bibitem{72} R Cawley and R. D. Mauldin, {\it Adv. in Maths} {\bf 92}, 192 (1992).
\bibitem{29} M. Holschneider {\it J. Stat. Phys.} {\bf 77}, 807 (1994). 
\bibitem{38} S. Jaffard, in {\it Wavelets and Applications} ed. Meyer Y. 
(Springer-Verlag, Berlin, 1992). 
\bibitem{25} A. Arneodo, E. Bacry and J. F. Muzy {\it Phys. Rev. Lett.} 
{\bf 74}, 4823 (1995).
\bibitem{51} S. Jaffard  and Y. Meyer, to appear in {\it Memoirs of the AMS}. 
\bibitem{47} S. Jaffard {\it Preprint}
\bibitem{48} S. Jaffard  and B. B. Mandelbrot, to appear in 
{\it Advances in Maths}.
\bibitem{49} I. Daubechies and J. Lagarias, {\it Rev. of Math. Phys.} {\bf 6},
1033 (1994).
\bibitem{52} S. Jaffard, {\it Preprint}
\bibitem{7}  K. B. Oldham  and J. Spanier   {\it The Fractional Calculus}
 ( Academic Press, New York, 1974)
\bibitem{8} K. S. Miller  and B. Ross  {\it An Introduction to the Fractional}
{\it Calculus and Fractional Differential Equations} 
( John Wiley, New York, 1993)
\bibitem{9} B. Ross, in {\it Fractional Calculus and its Applications: Lecture}
{\it Notes in    Mathematics}  ( Springer, New York, 1975) vol. 457 p. 1.
\bibitem{41} K. M. Kolwankar and A. D. Gangal {\it Chaos} {\bf 6}, 505 (1996).
(chao-dyn/9609016)
\bibitem{17} R. Hilfer  {\it Phys. Scr.} {\bf 44}, 321 (1991).
\bibitem{18} R. Hilfer  {\it Phys. Rev. Lett.} {\bf 68}, 190 (1992).
\bibitem{70} K. M. Kolwankar, {\it Preprint}
\bibitem{3} T. F. Nonnenmacher  {\it J. Phys. A: Math. Gen.}
 {\bf 23}, L697 (1990) 
\bibitem{4} M. Giona  and H. E. Roman  {\it J. Phys. A: Math Gen.} {\bf 25}, 
2093 (1992) 
\bibitem{5} H. E. Roman  and M. Giona   {\it J. Phys. A: Math. Gen.} {\bf 25},
2107 (1992)
\bibitem{6} N. Patzschke  and M. Z\"ahle  {\it Stochastic Process Appl.} 
{\bf 43}, 165 (1992)
\bibitem{11} B. B. Mandelbrot and J. W. Van Ness  {\it SIAM Rev.} {\bf 10}, 422 
(1968).
\bibitem{22} W. G. Gl\"ockle and T. F. Nonnenmacher {\it J. Stat. Phys.} 
{\bf 71}, 741 (1993).
\bibitem{23} M. F. Schlesinger {\it J. Stat. Phys.} {\bf 36}, 639 (1984).
\bibitem{55} K. L. Sebastian {\it J. phys. A: Math. Gen.} {\bf 28}, 4305 (1995). \bibitem{64} J. P. Bouchaud and A. Georges  {\it Phys. Rep.} 
{\bf 195}, 127 (1990).
\bibitem{65} M. F. Shlesinger, B. J. West and J. Klafter {\it Phys. Rev. Lett.}
{\bf 58}, 1100 (1987).
\bibitem{66} M. F. Shlesinger, G. M. Zaslavsky and J. Klafter {\it Nature}
 (London) {\bf 363}, 31 (1993).
\bibitem{71}M. F. Shlesinger et. al., Eds, 
{\it L\'evy Flights and Related Topics in Physics} (Springer, Berlin, 1995).
\bibitem{67} T. H. Solomon, E. R. Weeks and H. L. Swinney {\it Phys. Rev. Lett.}
{\bf 71}, 3975 (1993).
\bibitem{68} A. Ott, J. P. Bouchaud, D. Langvin and W. Urbach 
{\it Phys. Rev. Lett.} {\bf 65}, 2201 (1990).
\bibitem{60} W. Wyss J. Math. Phys. {\bf 27}, 2782 (1986).
\bibitem{53} W. R. Schneider and W. Wyss {\it J. Math. Phys.} {\bf 30}, 134 (1989).
\bibitem{54} G. Jumarie {\it J. Math. Phys.} {\bf 33}, 3536 (1992).
\bibitem{63} H. C. Fogedby {\it Phys. Rev. Lett.}{\bf 73}, 2517 (1994).
\bibitem{46} G. M. Zaslavsky {\it Physica D} {\bf 76}, 110 (1994).
\bibitem{42} A. Zygmund {\it Trignometric series} Vols I, II, 2nd ed.,
Cambridge Univ. Press, New York, 1959. 
\bibitem{43} E. M. Stein {\it Singular integrals and differentiability properties of functions} (Princeton University press, Princeton, 1970)
\bibitem{44} E. M. Stein and A. Zygmund {\it Proc. London Math. Soc.} (3)
{\bf 14a} 249 (1965).
\bibitem{45} G. V. Welland {\it Proc. Amer. Math. Soc.} {\bf 19}, 135(1968).
\bibitem{50} I. Eyink {\it J. Stat. Phys.} {\bf 78}, 353 (1995).
\bibitem{56} S. Jaffard, To appear in {\it SIAM J. of Math. Anal.}.
\bibitem{21} T. J. Osler {\it SIAM J. Math. Anal.} {\bf 2}, 37 (1971).
\bibitem{73} R. D. Mauldin and S. C. Williams, {\it Trans. Am. Math. Soc.}
{\bf 298} (2), 793 (1986).
\bibitem{79} B. R. Hunt, Preprint (1996).
\bibitem{12} G. H. Hardy  {\it Tr. Am. Math. Soc.} {\bf 17}, 301 (1916).
\bibitem{13} A. S. Besicovitch and H. D. Ursell {\it J. Lond. Math. Soc.} 
{\bf 12}, 18 (1937).
\bibitem{14} M. V. Berry and Z. V. Luwis {\it Proc. R. Soc.}{\bf A370}, 459 
(1980). 
\bibitem{16} M. F. Shlesinger {\it Physica D} {\bf 38}, 304 (1989). 
\bibitem{28} R. Benzi et. al. {\it Physica D}, {\bf 65}, 352 (1993). 
\bibitem{80} Juneja et. al., {\it Phys. Rev. E} {\bf 41} 5179 (1994).
\bibitem{57} G. H. Hardy and E. Littlewood {\it Acta Math.} {\bf 37}, 194 (1914).
\bibitem{58} J. Gerver {\it Am. J. Math.} {\bf 93}, 33 (1970).
\bibitem{61} M. Holschneider and Ph. Tchamitchian {\it Invent. Math.} {\bf 105}, 157
(1991).
\bibitem{62} J. J. Duistermaat, {\it Overdruk} {\bf 9}, 303 (1991).
\bibitem{75} L. Olsen, {\it Pitman Research Notes in Mathematica Series}
{\bf 307} (1994).
\bibitem{76} R. Riedi, {\it J. Math. Anal. Appl.} {\bf 189} 462 (1995).
\bibitem{77} M. Z\"ahle, Friedrich-Schiller Universit\"at Jena, Preprint (1996).
\bibitem{78} M. Z\"ahle, {\it Fractals} {\bf 3} 747 (1995).

\end{thebibliography}
\end{document}